\documentclass[runningheads]{llncs}

\usepackage{amsmath}
\usepackage{amssymb}
\usepackage{booktabs}
\usepackage[hyphens]{url}
\usepackage[hidelinks]{hyperref}

\newcommand{\fl}{\operatorname{fl}}
\newcommand{\RNE}{\operatorname{RNE}}
\newcommand{\rD}{r_{\mathrm{D}}}
\newcommand{\rH}{r_{\mathrm{H}}}
\newcommand{\reduce}{\operatorname{reduce}}
\newcommand{\unsat}{\textsc{unsat}}
\newcommand{\sat}{\textsc{sat}}
\newcommand{\qffp}{\textsf{QF\_FP}}
\newcommand{\qflia}{\textsf{QF\_LIA}}
\newcommand{\Bhat}{\widehat{B}}

\begin{document}

\title{Deciding Scale Placement in MXFP4 Dot Products with Integer Arithmetic }
\titlerunning{Deciding Scale Placement in MXFP4 Dot Products with Integer Arithmetic}
\author{Shreyas Vinayakumar}
\authorrunning{S. Vinayakumar}
\institute{}

\maketitle

\begin{abstract}
Block-scaled formats represent a group of values as low-bit elements
multiplied by one shared scale.  The Open Compute Project's microscaling
(MX) family includes MXFP4, which stores 32 four-bit elements with a shared
power-of-two scale.  While lowering a dot-product reduction for execution,
a compiler can apply these scales before the reduction, or defer scaling until afterwards.
Although equivalent over real numbers, these formulations can produce
different IEEE binary32 results through overflow or underflow, with
equivalence also depending on the reduction tree. We study the legality of compiler 
rewrites in either direction by deriving conditions under
which the two formulations are bitwise equivalent.

To decide rewrite legality, we construct an integer state machine
that exploits the finite element alphabet and shared power-of-two
scales, modeling overflow, NaN, and subnormal rounding.
We prove its faithfulness under bounded scale codes and integer
coefficients, yielding an exact encoding of the equivalence query
in quantifier-free linear integer arithmetic(\qflia).
For MXFP4, we derive a closed-form sufficient upper bound on the sum
of scale exponents for any fixed binary reduction tree within the supported
coefficient range. This bound depends on the size of the largest proper subtree.
For four tree shapes at block lengths $k\in\{2,3,4,8,16,32\}$ and
two shapes at $k=64$, SMT finds counterexamples when this upper
bound is increased by one exponent step.
We evaluate the computational cost of this approach against the standard direct encoding in quantifier-free floating-point
arithmetic (\qffp{}). On matched queries, cvc5 proves the integer version in seconds but reaches
a configured one-CPU-hour limit on the floating-point proof even at $k=2$.
The implementation was cross-checked using an
independent floating-point encoding, exhaustive and stratified binary32 execution, and
GPU replay of counterexamples.  An MLIR lowering prototype demonstrates
selection between the two formulations using the certified conditions as
static or runtime guards.

\end{abstract}

\section{Introduction}
\label{sec:intro}

As ML models, particularly large language models, continue to grow in size,
they require substantial memory and computational resources to store and process their parameters
and intermediate activations. Reducing the number of bits used to represent each value
can improve throughput and reduce energy use on supporting hardware compared to full
32-bit floating-point arithmetic, by reducing data movement and enabling efficient arithmetic.
However, reducing bit width affects precision and dynamic range,
and requires care to preserve numerical accuracy and to avoid overflow or underflow.

Block scaling formats like Open Compute Project's microscaling (MX) formats, which
store a tensor as blocks of narrow elements that share one scale~\cite{ocp2023mx,rouhani2023microscaling},
extend the dynamic range of the elements while keeping storage costs moderate.
These formats have become so important that major vendors
have added native support: NVIDIA's Blackwell tensor cores and AMD's matrix cores support multiplying 
block-scaled matrices~\cite{nvidia-cutlass,colfax2025blockscaling}.
On the software end, frameworks like Triton expose the operation as \texttt{dot\_scaled}
and emulate it on other targets~\cite{triton-dotscaled}, and \texttt{torchao} emulates it by
dequantizing to a wide type and calling an ordinary matrix
multiply~\cite{torchao}.

In MXFP4, a block holds 32 four-bit E2M1(2 bit exponent, 1 bit mantissa) elements and one eight-bit E8M0
scale, and the scale is a power of two. For an E2M1 element with sign bit $S$, exponent field $E$, and mantissa bit $M$,
its nonzero normalized decoded value is
\[
  (-1)^S 2^{E-1}\left(1 + \frac{M}{2}\right),
\]
which is then multiplied by the block scale. An E8M0 scale has the decoded value $2^{e-127}$,
where $e$ is the unsigned integer stored in the scale byte.
Consider the dot product of two k-sized MXFP4 blocks with scales $s_A$, $s_B$ and
elements $a_1,\dots,a_k$ and $b_1,\dots,b_k$.  A way to compute the dot product is to dequantize
elements from each block, and perform a reduction as shown below:
\[
  \sum_{i=1}^{k}(s_A a_i)(s_B b_i),
\]
Another way to compute it would be to perform reduction over the original element values without dequantization,
and then multiply the result with scales of both blocks:
\[
  (s_A s_B)\sum_{i=1}^{k}a_i b_i .
\]
We call these the \emph{direct} and
\emph{hoisted} forms respectively,
because it hoists the common scale for each element of a block.
The hoisted form performs $k+2$ multiplications instead of $3k$; at $k=32$ that is
34 instead of 96.  Both scale placements occur in practice: the OCP specification
uses the hoisted form while writing the block dot product but leaves the internal
precision and the order of operations to the implementation~\cite{ocp2023mx}.
On the other hand, torchao's emulation dequantizes operands before
matrix multiplication~\cite{torchao}.
Under real arithmetic, the distributive law makes the two forms equivalent.
This identity motivates scale hoisting as a compiler optimization.

In binary32 arithmetic, rewriting can change the answer significantly.
Take $k=32$, scale codes $e_A=120$ and $e_B=252$, which
denote $s_A=2^{-7}$ and $s_B=2^{125}$, and a block with twenty-nine lanes of
$a_i=b_i=6$ followed by the pairs $(-6,6)$, $(-2,2)$, and $(0,0)$.  The exact
dot product is $1004\cdot2^{118}\approx3.34\times10^{38}$.  This is below the
binary32 overflow threshold, and the hoisted form returns it.  The direct
form, summed from left to right, yields $29\cdot36\cdot2^{118}=4176\cdot2^{116}$
after twenty-nine lanes; that exceeds $2^{128}$, and the running sum overflows to
$+\infty$ before the remaining terms can be added.  The direct
form returns $+\infty$.  Summed as a balanced tree instead, the direct form
never builds that prefix and returns the finite value.  One input has
three plausible lowerings, and whether the results agree depends on the shape
of the reduction.  Table~\ref{tab:intro-counterexample} makes the resulting
failure of rewrite equivalence explicit.  Section~\ref{sec:mlir} shows the
direct and hoisted forms as an MLIR pass emits them.

\begin{table}[t]
  \centering
  \caption{The same 32-lane input evaluated by three plausible lowerings.
  The linear direct reduction is not equivalent to the hoisted form, so
  scale hoisting is not a legal result-preserving rewrite for this input and
  reduction tree.}
  \label{tab:intro-counterexample}
  \begin{tabular}{@{}lll@{}}
    \toprule
    Form & Reduction shape & Binary32 result \\
    \midrule
    direct  & linear (left fold) & $+\infty$ (\texttt{0x7f800000}) \\
    direct  & balanced tree     & $1004\cdot2^{118}$ (\texttt{0x7f7b0000}) \\
    hoisted & either shape      & $1004\cdot2^{118}$ (\texttt{0x7f7b0000}) \\
    \bottomrule
  \end{tabular}
\end{table}

Existing compilers have two ways to deal with this today.  Under strict IEEE semantics,
rewriting is forbidden, because it can change the returned value.  Under
fast-math flags it is allowed, together with every other reassociation. For example,
LLVM's language reference warns that reassociation may ``dramatically change
results''~\cite{llvm-langref}.  Neither option uses the unique constraints known
about this particular operation: the element alphabet has sixteen codes(4 bits per element), the
scale is a power of two, the block length is fixed by the format, and the
compiler itself chooses the reduction tree.  We derive sufficient guards for
rewrites between the direct and hoisted forms under a fixed reduction tree,
and establish sharpness for the configurations evaluated here.

Bit-precise SMT supports bounded equivalence checking of floating-point
programs~\cite{notzli2016lifejacket,lopes2021alive2}, but the direct encoding
is expensive even for small instances of this family
(Section~\ref{sec:qffp}).  Our contribution is an exact reduction to
quantifier-free linear integer arithmetic(\qflia{})
that exposes the arithmetic structure of these queries.

Every E2M1 product is a quarter-integer of
magnitude at most 36, and the scales are powers of two. Under an upper bound on
each scale code and a bound on the magnitudes of intermediate integer
coefficients, both programs can be represented by a state machine consisting of
an integer coefficient and a tag in $\{\mathrm{fin},+\infty,-\infty,\mathrm{NaN}\}$.
These bounds ensure exactness in the normal range, while overflow
and subnormal rounding can be expressed as conditions on integers.
The resulting query in \qflia{} is
equisatisfiable with the floating-point query under these bounds,
and counterexamples translate constructively in both directions
(Theorem~\ref{thm:faithful}, Section~\ref{sec:semantics}).

We use the following vocabulary throughout:  A \emph{lane} is an index
$i\in\{1,\dots,k\}$, one position in a block.  A \emph{scale code} is the
eight-bit E8M0 field $e$ of a block, and its scale is $2^{e-127}$.  The
\emph{exponent sum} $t=(e_A-127)+(e_B-127)$ is the exponent of the combined
scale $s_As_B$.  A \emph{band} is an interval of exponent sums.  The
\emph{guard} is the condition on the pair of scale codes under which we prove
the rewrite exact; it consists of an upper cap on each code and a band on the
exponent sum.  A band is \emph{certified} for a block length and a tree when
the solver has proved that no block inside it separates the two forms, and it
is \emph{sharp} when widening it by one exponent step admits a
counterexample.  A \emph{witness} is a concrete block, elements and scale
codes together, on which the two forms return different binary32 values.

This paper makes the following contributions.
\begin{itemize}
\item An exact reduction of a family of floating-point equivalence queries to
  \qflia{}, with a faithfulness proof and constructive counterexample
  translations on the domain of interest (Section~\ref{sec:semantics}).
\item A closed-form sufficient guard using MXFP4's element and scale encodings
  for any fixed binary reduction tree
  within the coefficient budget, with solver-established sharpness for
  four shapes at block lengths 2,3,4,8,16,32 lanes and for two shapes at 64 lanes
  (Sections~\ref{sec:bands} and~\ref{sec:sharpness}).
\item A prototype implementation of a rewrite guard based on the certified
bands (Section~\ref{sec:mlir}) in an MLIR lowering pass, and 
a description of which parts of the construction carry over to
  other block formats(Section~\ref{sec:compiler}).
\item A comparison of the two theories,
  evaluating the relative cost using the same solver and 
  validating the encoding against an independently written floating-point
  encoding, against exhaustive and stratified binary32 execution, and by
  replaying witnesses on a GPU(Section~\ref{sec:eval}).
\end{itemize}

\section{The Rewrite and Its Legality Question}
\label{sec:problem}

\subsection{Formats}

An E2M1 element has a sign bit, two exponent bits, and one mantissa bit.  Its
sixteen codes denote $0$, $\pm0.5$, $\pm1$, $\pm1.5$, $\pm2$, $\pm3$, $\pm4$,
and $\pm6$, with separate codes for $+0$ and $-0$; there are no codes for
infinity or NaN.  We write an element as $a=m/2$ with
\[
  m\in M=\{0,\pm1,\pm2,\pm3,\pm4,\pm6,\pm8,\pm12\},
\]
and note that every nonzero $|m|$ has at most two significant bits.

An E8M0 scale is an eight-bit exponent without sign or mantissa.  Code
$e\in[0,254]$ denotes $2^{e-127}$, so the scales run from $2^{-127}$ to
$2^{127}$; code 255 is NaN.  We denote $x=e_A-127$ and $y=e_B-127$ for the two
unbiased exponents, which implies $s_A=2^x$ and $s_B=2^y$, and $t=x+y$ for the
exponent sum.  The OCP block length for MXFP4 is $k=32$~\cite{ocp2023mx}.

\subsection{Binary32 arithmetic}

All arithmetic is IEEE binary32 with round to nearest, ties to even, and
gradual underflow~\cite{ieee754}.  We denote $\fl(v)$ for the correctly rounded value of a
real number $v$; every operation returns $\fl$ of its exact result.  We use
the following facts:
The smallest positive value is $\delta=2^{-149}$, and every subnormal value is
an integer multiple of $\delta$.  A real $v$ is representable,
and hence $\fl(v)=v$, if $v=0$, or if
$|v|<2^{-126}$ and $v$ is a multiple of $\delta$, or if
$2^{-126}\le|v|<2^{128}$ and $v$ has at most 24 significant bits.
Infinity and NaN follow IEEE rules: $\infty+(-\infty)$ and $0\cdot\infty$ produce NaN,
an infinity absorbs any finite value, and NaN propagates through every operation.

\subsection{The two programs}

A reduction tree $\rho$ over $k$ lanes is a fixed binary tree with the lanes
as leaves; each internal node performs one binary32 addition of its two
children.  For a tree $\rho$, the direct form computes
\begin{align}
  p_i &= \fl\big(\fl(s_A a_i)\cdot\fl(s_B b_i)\big), &
  \rD^\rho &= \reduce_\rho(p_1,\dots,p_k), \label{eq:direct}
\end{align}
and the hoisted form computes
\begin{align}
  q_i &= \fl(a_i b_i), &
  \rH^\rho &= \fl\big(\fl(s_A s_B)\cdot\reduce_\rho(q_1,\dots,q_k)\big).
  \label{eq:hoisted}
\end{align}
Each multiplication and addition is a separate rounded operation; we do not
model fused multiply-add.  We consider four trees for the reduce operation in (1) and (2): the \emph{left fold},
which adds lanes in order into a running sum; the \emph{right fold}, which
does the same starting from the last lane; the \emph{balanced tree}, which
halves the lane sequence recursively; and a \emph{quarter-skew tree}, which
places $\max(1,\lfloor n/4\rfloor)$ of an $n$-lane node's lanes in its left
child and splits both children recursively.  The last shape is there to test
a tree that is neither a chain nor balanced.

\subsection{The equivalence relation}

We compare results with a relation $\simeq$ under which $+0\simeq-0$, any two
NaNs are equal, and finite values and infinities must match exactly.
In SMT-LIB this is expressed by \texttt{fp.eq} augmented with a test
that both operands are NaN~\cite{smtlib-fp}.
A checker for scalar IEEE optimizations must also distinguish the sign of
zero~\cite{notzli2016lifejacket}; we write $\equiv$ for that stricter
relation.  We state results for $\simeq$ and show in
Section~\ref{sec:signed-zero} that the two relations agree on the certified
domain, so the choice does not affect the guard.

\subsection{The guard}
\label{sec:guard}

Without restrictions the two forms are not equivalent, and the solver finds
several kinds of witness: either form can overflow to $\pm\infty$ while the other is
finite, as in Section~\ref{sec:intro}; either form can produce NaN (from
$\infty-\infty$ or from $0\cdot\infty$) while the other is finite;
the direct form can accumulate subnormal
values while the hoisted form's combined scale $\fl(s_As_B)$ underflows to zero.
Every witness found pushes a value to the end of the binary32 range.
We seek a scale-only condition $P$ on the two scale codes such that, for $k\ge2$,
\begin{equation}
  \forall a,b,e_A,e_B.\quad P(e_A,e_B)\implies\rD^\rho\simeq\rH^\rho .
  \label{eq:query}
\end{equation}
We consider guards $P_B$ consisting of two parts: individual bounds on the scale
codes and an interval constraint on their exponent sum.
The examples below show how equivalence can fail when these restrictions are relaxed.
We subsequently establish sufficiency and determine the largest admissible upper endpoint.

We impose the following individual bounds on scale codes:
\begin{equation}
  0\le e_A,\,e_B\le252 . \label{eq:cap}
\end{equation}
Codes 253 and 254 are excluded because each breaks the rewrite on its own.
With $e_A=254$, $e_B=0$, and $(a_1,b_1)=(6,0.5)$, the direct form computes
$\fl(s_Aa_1)=\fl(6\cdot2^{127})=+\infty$ in its first lane, while the hoisted
form computes $\fl(2^0\cdot3)=3$; code 253 fails the same way one exponent
lower.  The bound is a property of the rewrite.  Code 0, representing $2^{-127}$,
is admitted, and code 255 is the NaN encoding which is
never admitted.  We use the term 
\emph{capped domain} to refer to the blocks whose codes satisfy \eqref{eq:cap}.

We impose this interval constraint on the exponent sum $t$:
\begin{equation}
  -147\le t\le B, \label{eq:band}
\end{equation}
with a parameter $B$ to be determined.  The lower endpoint is based on
underflow. At $t=-148$, two lanes with $a_i=b_i=0.5$ yield
$p_i=\fl(2^{-150})=0$ for the direct form,
since $2^{-150}$ lies halfway between $0$ and $\delta$
and ties go to the even side, while the hoisted form yields
$\fl(2^{-148}\cdot0.5)=\delta$ exactly.
Next, we determine the largest admissible upper endpoint.
Let $B^{\max}_\rho(k)$ be the largest $B$ for which \eqref{eq:query}
holds with the guard \eqref{eq:cap}, \eqref{eq:band}.
Section~\ref{sec:bands} derives a closed-form sufficient endpoint
$\Bhat_\rho(k)\le B^{\max}_\rho(k)$, and the solver establishes equality
for the configurations analyzed in Section~\ref{sec:eval}.

In terms of the eight-bit codes the guard can be written as
\begin{equation}
  e_A\le252,\qquad e_B\le252,\qquad 107\le e_A+e_B\le254+B,
  \label{eq:encoded}
\end{equation}
four comparisons on two bytes.  This is what a compiler needs to evaluate, statically
or at run time.

\subsection{Floating-point SMT as a baseline}
\label{sec:qffp}

The negation of query \eqref{eq:query} for a fixed $k$ is a quantifier-free network of 
binary32 operations with bounded elements
and codes.  We encoded it directly in SMT-LIB
floating-point theory (\qffp, using \textsf{QF\_BVFP} for the bit-vector inputs).

To isolate the effect of theory choice from solver choice, we use paired left-fold
queries with identical domains: $1\le e_A,e_B\le252$ and $-B\le t\le B$.
cvc5~\cite{barbosa2022cvc5} proves the two- and four-lane integer queries in under one second
on average, but the corresponding floating-point proofs reach the
configured one-CPU-hour cap.  Among the FP solvers tested, Bitwuzla is
fastest on the completed safe-boundary queries: it proves four lanes in
499 CPU seconds and completes eight lanes after at least 69\,238 CPU seconds,
as bounded by the last progress sample.  At eight lanes, cvc5's integer query averages
4.78 wall-clock seconds.  These comparison queries use a narrower domain
than the full guard certified in Section~\ref{sec:eval}.

\begin{table}[t]
  \caption{Matched left-fold safe-boundary queries on the comparison domain
    ($1\le e_A,e_B\le252$, $|t|\le B$). LIA timings are average wall-clock seconds over
   three runs; FP timings are CPU seconds from one run.
    ``cap'' denotes termination at the configured 3600-second CPU limit;
    the eight-lane Bitwuzla query returned \unsat{}, with CPU time bounded
    below by the last progress sample before completion.
    A dash denotes an unmeasured query.}
  \label{tab:qffp}
  \centering
  % Generated by analyze_theory.py from frozen CSV and solver logs.
\begin{tabular}{@{}rr rrrr@{}}
\toprule
$k$ & $B$ & cvc5 LIA & cvc5 FP & Bitwuzla FP & Z3 FP \\
\midrule
2 & 122 & 0.46 & cap & 14.10 & 1251.02 \\
4 & 121 & 0.89 & cap & 499.10 & --- \\
8 & 120 & 4.78 & --- & $\ge69\,238$ & --- \\
\bottomrule
\end{tabular}

\end{table}

\section{An Exact Integer Semantics}
\label{sec:semantics}

The floating-point encoding asks the solver to reason about 32 pairs of
four-bit codes, two scale codes, and a network of roughly a hundred binary32
operations, each of which can in principle round.  On the capped domain
very few of them do.  We identify a single power-of-two grid
on which every value of both programs lies,
reduce each operation to integer arithmetic on that grid plus an overflow test and a
subnormal rounding step, and prove that the reduced semantics computes
exactly what binary32 computes.

\subsection{One shared grid}

Let
\[
  n_i=m_{a,i}\,m_{b,i},\qquad |n_i|\le144,\qquad E=t-2 .
\]
Exact raw products in the hoisted form are of the form $q_i=n_i/4$.  They are multiples
of $2^{-2}$ with a coefficient below $2^8$, representable in binary32. 
Under the coefficient-budget assumption $144k\le2^{24}$,
any partial sum of raw products is $N/4$ with $|N|\le144k\le2^{24}$.
It is exactly representable, so the hoisted raw reduction is exact for any fixed tree.

Under \eqref{eq:cap}, each scaled element of the direct form
is exactly representable in binary32 as a finite value:
\begin{equation}
  \fl(2^x a_i)=m_{a,i}\,2^{x-1}. \label{eq:single-scale}
\end{equation}
The magnitude is at most $12\cdot2^{124}$,
below the largest finite value $\Omega=(2^{24}-1)\,2^{104}$;
in the normal range the coefficient has at most two significant bits; and in the subnormal
range the value is a multiple of $2^{x-1}\ge2^{-128}$, hence of $\delta$.
Individual scaling is the step that requires the cap, since codes 253 and 254 would overflow without it.
The exact real product of the scaled elements is $n_i2^{E}$, so the direct leaf obtained after rounding is
\begin{equation}
  p_i=\fl(n_i\,2^{E}). \label{eq:leaf}
\end{equation}
Every finite intermediate of the direct form is therefore an integer multiple
of $2^E$, or, after rounding into the subnormal range, an integer multiple of
$\delta$.  The whole computation of both programs therefore takes place on one grid whose
spacing $2^E$ is determined by the two scale codes.

\subsection{The rounding primitive}
\label{sec:primitive}

The next step is to demonstrate that $\fl$ operating on a grid value still produces representable values.
For an integer $n$
with $|n|\le2^{24}$ and $E\in[-256,248]$, the range the capped domain produces,
\begin{equation}
\fl(n\,2^E)=
\begin{cases}
 \operatorname{sign}(n)\cdot\infty, & |n|\,2^E\ge2^{128},\\
 n\,2^E, & E\ge-149\text{ and no overflow},\\
 \RNE(n/2^d)\,\delta, & E\le-150,\ d=-149-E .
\end{cases}
\label{eq:rounding}
\end{equation}
Here $\RNE(n/2^d)$ is the integer nearest to $n/2^d$, with ties to the even
integer.  The middle case is exactness: a multiple of $2^E$ with a
coefficient of magnitude at most $2^{24}$ has at most 24 significant bits, and
when $E\ge-149$ it is a multiple of $\delta$, so it is representable whether it
lands in the normal or the subnormal range.  The last case is subnormal
rounding: for $E\le-150$ the magnitude is at most $2^{24}\cdot2^{-150}=2^{-126}$,
every representable value below $2^{-125}$ is a multiple of $\delta$, and IEEE
rounding onto the uniform grid $\delta\mathbb{Z}$ with ties to even is the
integer shift.

The first case requires a separate argument, because binary32 does not overflow at $2^{128}$.
It overflows at $\Omega^+=2^{128}-2^{103}$: a value in $(\Omega,\Omega^+)$
rounds down to $\Omega$, and a value in $[\Omega^+,2^{128})$ rounds up to
$\infty$.  If a grid value could land in $(\Omega,2^{128})$, the integer test
in \eqref{eq:rounding} would be wrong on one side or the other.  None can.  A
grid value can reach $\Omega$ only if $2^E\ge\Omega/2^{24}>2^{103}$, so
$E\ge104$; and for $E\ge104$ the multiples of $2^E$ skip the whole interval
$(\Omega,2^{128})$, whose width $2^{104}$ does not exceed their spacing, the
largest multiple below $2^{128}$ being $2^{128}-2^E\le\Omega$.  The simpler
test $|n|2^E\ge2^{128}$ is therefore exact on this domain.  At $k\le32$ the coefficient
bound $4608$ tightens the threshold to $E\ge116$, which is the window the
encoding expands over; in general overflow requires
$E\ge128-\lfloor\log_2144k\rfloor$, and Section~\ref{sec:cost} shows the failure 
mode at $k=64$ if an encoding keeps the 32-lane window.

\subsection{The state machine}

For $E\ge-149$ we represent the direct reduction by a state
\[
  (\mathit{kind},S),\qquad
  \mathit{kind}\in\{\mathrm{fin},+\infty,-\infty,\mathrm{NaN}\},
\]
where a finite state denotes the value $S\,2^E$.  A leaf starts as
$(\mathrm{fin},n_i)$ unless $|n_i|2^E\ge2^{128}$, in which case it is
$\pm\infty$.  Adding two finite states adds their integers and applies the
overflow test of \eqref{eq:rounding} to the sum.  Every other combination
follows the IEEE rules: a finite state and an infinity yield the infinity,
opposite infinities give NaN, and NaN is absorbing.  There is no other source
of NaN in the direct form, because by \eqref{eq:single-scale} every
multiplication has two finite exact operands.

For $E\le-150$ every leaf rounds to $\RNE(n_i/2^d)\delta$,
 where $d=-149-E\ge1$. The partial sums are then multiples of
$\delta$ with coefficient magnitude at most $72k$. Under $144k\le2^{24}$,
every partial is exactly representable. Even if the leaves round,
every addition is exact including at the smallest normal value.

The hoisted form needs the same primitive twice.  The combined scale is
$\fl(2^x2^y)=\fl(2^t)$, which is $2^t$ for $-149\le t\le127$, $0$ for
$t\le-150$, and $+\infty$ for $t\ge128$.  The raw reduction is exact, as
shown above, and the final multiplication is $\fl(N\,2^E)$, which is
\eqref{eq:rounding} again, with the two special cases
$\infty\cdot0=\mathrm{NaN}$ and $0\cdot N=\pm0$ when the scale is infinite or
zero.

\subsection{Faithfulness}
\label{sec:faithful}

The argument above is organized as six lemmas.  Each has a short proof from
the representability characterization and the RNE overflow threshold, and
each finite-domain lemma is also checked exhaustively against
binary32 (Section~\ref{sec:validation}).

\begin{description}
\item[L1, single-scale exactness.] For $x\in[-127,125]$ and every E2M1 code,
  $\fl(2^xa_i)$ equals $m_{a,i}2^{x-1}$ exactly and is finite.
\item[L2, rounding primitive.] For $|n|\le2^{24}$ and $E\in[-256,248]$,
  \eqref{eq:rounding} computes $\fl(n2^E)$.
\item[L3, grid addition.] If $u=s_12^g$ and $w=s_22^g$ are binary32 values
  with $g\ge-149$ and $|s_1+s_2|\le2^{24}$, then $\fl(u+w)=(s_1+s_2)2^g$ unless
  $|s_1+s_2|2^g\ge2^{128}$, in which case it is $\pm\infty$ by the sign of
  the sum.
\item[L4, the overflow gap.] No value $n2^E$ with integer $1\le n\le2^{24}$
  lies in $(\Omega,\Omega^+)$, and none lies in $[\Omega^+,2^{128})$.
\item[L5, raw products.] $\fl(a_ib_i)=n_i/4$ for all 256 code pairs;
  if $144k\le2^{24}$, every partial sum equals $N/4$ exactly, for any tree.
\item[L6, the scale product.] $\fl(2^x2^y)=2^t$ for $-149\le t\le127$, $0$
  for $t\le-150$, and $+\infty$ for $t\ge128$.
\end{description}

\begin{theorem}[Faithfulness]
\label{thm:faithful}
Let $k$ satisfy $144k\le2^{24}$, let the scale codes satisfy \eqref{eq:cap},
let all operations be binary32 with round to nearest even and gradual
underflow, and let $\rho$ be any fixed binary reduction tree.  Then the
integer state machine computes a value equivalent under $\simeq$ to
the direct form \eqref{eq:direct} and, separately, to the hoisted form
\eqref{eq:hoisted}.
\end{theorem}

\begin{proof}
For the direct form with $E\ge-149$, L1 makes both
factors of every leaf exact, so the leaf is $\fl(n_i2^E)$, and L2 gives its
state.  By induction over the tree, every finite node holds an exact multiple
$S2^E$ with $|S|\le144k$, and L3 says its addition is exact unless the
integer sum overflows, in which case the sign of the sum selects the
infinity.  The infinity and NaN cases follow from the IEEE rules.  For
$E\le-150$, L2 rounds each leaf onto the $\delta$ grid and L3 with $g=-149$
makes every later addition exact.  For the hoisted form, L6 gives the
combined scale, L5 the exact raw reduction, and L2 the final multiplication;
the cases of an infinite or zero scale are again the IEEE rules.  Since the
machine preserves both results under $\simeq$, its disagreement predicate
and the floating-point predicate have the same truth value.
\end{proof}

The bound $144k\le2^{24}$, that is $k\le116\,508$, is required for representability
of partial sums in L3 and L5. It is far above block lengths specified by the format.
The theorem establishes that the integer state machine captures the floating-point behavior up to $\simeq$ under the stated assumptions.
An \unsat{} proof of the integer query is a proof of the floating-point statement,
and a satisfying assignment of the integer query yields a concrete witness.
Each concrete block determines $n_i=m_{a,i}m_{b,i}$ and $t=x+y$;
conversely, every integer in the 37-element product alphabet factors into two E2M1 codes,
and every $t\in[-254,250]$ splits into two capped exponents. 
We validate each reported witness by reconstructing the E2M1 and E8M0 codes
and replaying computation in binary32.

\subsection{The integer query}
\label{sec:query}

Our integer encoder turns the state machine into a \qflia{} formula for Z3.  Each
lane is represented by one integer variable $n_i$ constrained to the 37 signed
values that arise as products of two elements of $M$.  One further integer
variable $t$ carries the exponent sum under the band \eqref{eq:band};
$E=t-2$ needs no variable of its own.  The formula for $k$ lanes has $k+3$
assertions: one alphabet constraint per lane, the two band bounds, and the
negation of $\rD^\rho\simeq\rH^\rho$.

The reduction tree is a function from a lane interval to two subintervals,
applied recursively, so a tree shape is a parameter of the encoding rather than a
separate encoding.  Each internal node combines two $(\mathit{kind},S)$
states.  The overflow test $|S|2^E\ge2^{128}$ is not linear in $E$, so the
encoding expands it over the finitely many exponents at which overflow is
possible, $E\ge128-\lfloor\log_2144k\rfloor$, as a nested conditional with
one integer comparison per exponent.  The subnormal branch is handled the
same way: for each shift $d$ from $1$ to $1+\lfloor\log_2n_{\max}\rfloor$,
where $n_{\max}$ is 144 at a direct leaf and $144k$ at the hoisted root,
division and remainder by the constant $2^d$ implement round to nearest even,
including the parity test on ties; for larger $d$ every coefficient rounds to
zero.  Division by a constant is linear, so the formula stays in \qflia{}.
The hoisted side is shared across tree shapes because its raw reduction is
exact and shape-invariant.

For a fixed product alphabet bounded by $C$ and a fixed exponent sum,
each node has $O(Ck)$ direct $(\mathit{kind},S)$ states.
In addition, raw sums contribute the same number giving $O((Ck)^2)$ joint states.
Bottom-up enumeration can be used to decide disagreement in polynomial time in $k$
for fixed $C$ and the finite set of exponents in the band. 
While QF\_LIA is NP-complete in general, the restricted query family
allows for polynomial-time decision procedures.

\section{Certified Bands}
\label{sec:bands}

Searching over $B$ with solver queries determines $B^{\max}_\rho(k)$.
We derive a sufficient upper endpoint from the maximum absolute
partial sum possible at a proper subtree of the reduction tree. This endpoint
matches the solver results for the tested configurations.

\subsection{A safe endpoint for any tree}

Let $m_\rho$ be the largest number of leaves below any proper (non-root) node for fixed $k\ge2$,
$m_\rho=1$ when $k=2$. Let $P_\rho=144m_\rho$ be the largest integer coefficient a proper node
can hold.  For a left or right fold $m_\rho=k-1$, for a balanced tree
$m_\rho=\lceil k/2\rceil$, and for the quarter-skew tree we read $m_\rho$ off
its recursive split.

Inside the band, the direct form and the hoisted form can differ only if
some node of the direct reduction overflows while the hoisted root does not,
or the other way around.  The hoisted root and the direct root see the same
integer, so they overflow together.  A proper node of the direct tree,
however, holds a partial sum that the hoisted form never rounds, and if that
partial sum overflows, the infinity persists to the root while the hoisted
form may still be finite.  Keeping every proper node finite at the top of
the band requires $P_\rho2^{B-2}<2^{128}$, that is $B<130-\log_2P_\rho$.
Since $P_\rho$ is a multiple of nine it is not a power of two, so the largest
admissible integer is
\begin{equation}
  \Bhat_\rho(k)=\lfloor130-\log_2P_\rho\rfloor=129-\lfloor\log_2P_\rho\rfloor .
  \label{eq:bhat}
\end{equation}

\begin{theorem}[Safe band]
\label{thm:safe}
Assume separate binary32 operations with round to nearest even and gradual
underflow. For $k\ge2$ with $144k\le2^{24}$, any fixed binary tree $\rho$,
and scale codes satisfying \eqref{eq:cap} and $-147\le t\le\Bhat_\rho(k)$,
$\rD^\rho\simeq\rH^\rho$.
\end{theorem}

\begin{proof}
Since $t\ge-147$, $E\ge-149$, so by L1 and L2 every direct leaf is the exact
grid value $n_i2^E$: no leaf overflows, because
$|n_i|2^E\le144\cdot2^{\Bhat-2}<2^{128}$, and no leaf is rounded, because
$E\ge-149$.  We show by induction over the tree that every proper node holds
the exact value $S2^E$ of its subset sum.  For a proper node with finite
exact children, L3 makes the addition exact unless $|S|2^E\ge2^{128}$; but
$|S|\le144m_\rho=P_\rho$ and $t\le\Bhat_\rho(k)$ give
$|S|2^E\le P_\rho2^{\Bhat-2}<2^{128}$.  At the root, L3 gives either the exact
sum $N2^E$ or $\pm\infty$ by the sign of $N$, according to whether
$|N|2^E\ge2^{128}$.  On the hoisted side, L6 gives the combined scale $2^t$
exactly, since $-147\le t\le\Bhat\le127$; L5 gives the raw sum $N/4$ exactly;
and the final multiplication is $\fl(N2^E)$, which by L2 is the exact value
or $\pm\infty$ under the same test with the same sign.  The two results are
equal, including when both are infinite.
\end{proof}

The above proof does not rely on a solver, and it holds for every binary tree, not
only the four we measure.

\subsection{The lower endpoint}

The lower endpoint $-147$ is independent of the reduction tree. For $t\ge-147$,
$E\ge-149$ satisfying the condition used in the previous proof.  At $t=-148$
the two-lane witness of
Section~\ref{sec:guard} establishes non-equivalence for any tree, since both lanes
are leaves and we can pad with zero lanes.
Consequently, the certified band cannot be extended below $-147$.

\subsection{Sharpness}

Theorem~\ref{thm:safe} establishes that the band is safe.  Its sharpness also depends on the element
alphabet. To establish sharpness at $\Bhat_\rho(k)+1$, a sufficient proof is to construct an input whose
coefficient magnitude reaches $2^{129-\Bhat}$ while later lanes or a sibling subtree cancel enough of it that the
root stays below the threshold. The construction must use only the 37 admissible product values.
Such examples were found for all tested block lengths and trees,
and Section~\ref{sec:sharpness} reports the solver's confirmation that $B^{\max}_\rho(k)=\Bhat_\rho(k)$ in
all 26 measured cases.
Sharpness for every binary tree remains unproven but holds for the ones we have checked.

\subsection{Signed zeros}
\label{sec:signed-zero}

Our relation $\simeq$ identifies $+0$ with $-0$.  Inside the certified band
the stricter relation $\equiv$ gives the same verdict: every direct leaf is
exactly $2^t$ times its hoisted leaf and has the same zero sign, grid
additions are exact and combine the same signed operands at corresponding
nodes, and the positive final scale preserves the sign of a zero root.  So
$\rD^\rho\simeq\rH^\rho$ if and only if $\rD^\rho\equiv\rH^\rho$ on the
certified domain, and nothing in the guard depends on the choice of
relation.

Below the band, the 2 forms differ in the signs of zero.  At $t=-151$, with
capped codes $e_A=1$, $e_B=102$, $a=(1,-2)$, and $b=(1,1)$, the direct form
returns $+0$ and the hoisted form $-0$.  Exhaustive enumeration at $k=2$
finds the largest such $t$ at $-148$, one step below the lower endpoint.

\subsection{Revisiting the example}

The block in Section~\ref{sec:intro} has $k=32$ and $t=118$.
For a left fold reduction tree, $m_\rho=31$ and $P_\rho=4464$, giving
$\lfloor\log_2 4464\rfloor=12$ and $\Bhat=117$.
It therefore lies one exponent step above the certified band.
Its 29-lane prefix has coefficient $4176>4096$, causing overflow
at a proper node of the reduction tree.

For the balanced tree, $m_\rho=16$ and $P_\rho=2304$, giving
$\Bhat=118$ for the block, placing it within the certified band.
The two 16-lane subtrees have raw sums $576$ and $428$,
corresponding to coefficients $2304$ and $1712$, both below
the overflow threshold $4096$.
Theorem~\ref{thm:safe} therefore guarantees that the balanced
direct form and the hoisted form return the same result,
consistent with the evaluation in Section~\ref{sec:intro}.

\section{Evaluation}
\label{sec:eval}

We evaluate the encoding's fidelity
(Section~\ref{sec:validation}), sharpness of endpoints (Section~\ref{sec:sharpness}),
certification cost (Section~\ref{sec:cost}),
and conformance with modeled hardware arithmetic (Section~\ref{sec:hardware}).

\subsection{Setup}

The matched-query comparison uses cvc5 1.4.1, Bitwuzla 0.9.1, and
Z3 5.0.0, with Python 3.12 and NumPy 2.5, on a four-vCPU WSL2 host with
about 8\,GiB of memory. The matched-query measurements used two concurrent
solver processes; the separate Z3 certification study reports idle-host wall times.
The integer encoder, the independent floating-point encoder, the
lemma checks, the differential tests, and the witness concretizer are Python
scripts of a few hundred lines each.

\subsection{Validation of the encoding}
\label{sec:validation}

We validate the implementation against the mathematical model and against binary32 execution at three
levels.

First, each finite-domain lemma is checked over its entire domain against
NumPy's binary32. For the rounding checks, the reference calculates $n2^E$
exactly in binary64 and converts it once to binary32 using round-to-nearest,
ties-to-even. The checks covered all $16\times253$
single-scale products (L1), all $9217\times505$ scaled-integer points (L2),
all 256 raw code-pair products (L5), all 505 exponent sums of the scale
product (L6), and the exact integers around the overflow gap (L4).  Grid
addition (L3) has a two-dimensional domain, so the analytical proofs were supplemented by
200\,000 tests in the region near the overflow boundary.

Second, the composed state machine is compared under $\simeq$ with binary32
execution of both programs: exhaustively for $k=1$, over every code pair and
every exponent sum, and on 140\,000 stratified random blocks for
$k\in\{2,3,4,8,32\}$, and a separate balanced-tree test covering $k=16$.
Sampling was done at code 0, in deep underflow, at the subnormal
transition, in the central exact range, at the overflow boundary, and at
scale-product overflow.  No mismatches were detected.

Third, the \qflia{} query and the independently written \qffp{} query were checked
for agreement on the same boundaries at $k=2$ and $k=4$.  They agree on all four
verdicts: satisfiable at $B=123$ and unsatisfiable at $B=122$ for two lanes,
satisfiable at $B=122$ and unsatisfiable at $B=121$ for four.  Every witness
the integer query reports was factored into E2M1 and E8M0 codes
and replayed in binary32 reproducing every reported divergence.

\subsection{Validating sharpness}
\label{sec:sharpness}

Table~\ref{tab:boundaries} lists $B^{\max}_\rho(k)$ for six block lengths
and four trees.  Each entry is a pair of solver runs: \unsat{} at the listed
$B$ and \sat{} at $B+1$.  For the larger satisfiable queries we pin $t=B+1$
instead of leaving it free in the band, which restricts the search to the edge
and does not weaken the certificate, since a witness at a single $t$ is a
witness for the band.  All 24 entries equal $\Bhat_\rho(k)$ from
\eqref{eq:bhat}, as do the two $k=64$ entries in Table~\ref{tab:scaling}.

\begin{table}[t]
  \caption{Largest certified endpoint $B^{\max}_\rho(k)$ under the cap
    $0\le e_A,e_B\le252$ and the band $-147\le t\le B$.  Every entry is
    \unsat{} at $B$ and \sat{} at $B+1$.}
  \label{tab:boundaries}
  \centering
  \begin{tabular}{@{}rcccc@{}}
    \toprule
    $k$ & left fold & balanced & right fold & quarter-skew \\
    \midrule
     2 & 122 & 122 & 122 & 122 \\
     3 & 121 & 121 & 121 & 121 \\
     4 & 121 & 121 & 121 & 121 \\
     8 & 120 & 120 & 120 & 120 \\
    16 & 118 & 119 & 118 & 119 \\
    32 & 117 & 118 & 117 & 118 \\
    \bottomrule
  \end{tabular}
\end{table}

The boundaries for the left and right folds are identical with
$m_\rho=k-1$.  The balanced tree minimizes the size of the largest
proper subtree with size $m_\rho=\lceil k/2\rceil$.
The quarter-skew tree has $m_\rho=1,2,3,6,12,24$
at the listed lengths, larger than the corresponding balanced values for $k\ge4$ but 
yielding the same value of 
$\lfloor\log_2(144m_\rho)\rfloor$ in \eqref{eq:bhat}. The boundaries are 
identical across all 4 shapes at tabulated lengths through $k=8$.
For $k=16$ and $k=32$, there is an additional admissible exponent
step for balanced and quarter-skew trees relative to folds.
At $k=32$, the balanced and quarter-skew tree guards accept
the block of Section~\ref{sec:intro}, while the fold guards do not.

\subsection{Certification time measurement}
\label{sec:cost}

Table~\ref{tab:scaling} reports Z3 \unsat{} proofs at each boundary for
left-fold and balanced trees, at block lengths from 2 to 64. Each query was
run in three fresh processes on an otherwise idle four-vCPU WSL2 guest;
CPU-load observations and all individual logs are included in the artifact.
Wall time includes formula construction, DAG counting and solving, and
excludes imports, process startup and log writing. Assertions are top-level
constraints; DAG size counts distinct Z3 AST nodes reachable from all
assertions before simplification, counting shared nodes once.

All 36 runs returned \unsat{}. Mean times range from 0.11 to 6.92 seconds.
At 32 lanes, the left fold averages 1.76 seconds and the balanced tree
2.36 seconds. Formula size grows with $k$; the two shapes have identical
node counts at each length because only their wiring differs. These
measurements report certification cost for this implementation and host;
three repetitions do not establish a general solver-scaling law.

\begin{table}[t]
  \caption{\qflia{} \unsat{} proofs at the certified endpoint, using the
    supplied encoding. Times are means and ranges of three wall-clock runs
    of Z3 on an otherwise idle host. The table is generated from the
    individual run records included in the artifact.}
  \label{tab:scaling}
  \centering
  \setlength{\tabcolsep}{5pt}
  % Generated by collect_table4.py from individual fresh run records.
\begin{tabular}{@{}lrrrrrr@{}}
\toprule
Tree & $k$ & $B^{\max}$ & Asserts & DAG & Time (s) & Range (s) \\
\midrule
left fold & 2 & 122 & 5 & 954 & 0.17 & 0.10--0.26 \\
left fold & 4 & 121 & 7 & 1629 & 0.76 & 0.68--0.86 \\
left fold & 8 & 120 & 11 & 2954 & 1.17 & 1.09--1.22 \\
left fold & 16 & 118 & 19 & 5579 & 1.74 & 1.67--1.82 \\
left fold & 32 & 117 & 35 & 10805 & 1.76 & 1.37--2.15 \\
left fold & 64 & 116 & 67 & 21489 & 3.12 & 2.49--3.64 \\
\addlinespace
balanced & 2 & 122 & 5 & 954 & 0.11 & 0.10--0.12 \\
balanced & 4 & 121 & 7 & 1629 & 0.71 & 0.55--0.84 \\
balanced & 8 & 120 & 11 & 2954 & 0.94 & 0.80--1.18 \\
balanced & 16 & 119 & 19 & 5579 & 1.40 & 1.25--1.68 \\
balanced & 32 & 118 & 35 & 10805 & 2.36 & 2.23--2.57 \\
balanced & 64 & 117 & 67 & 21489 & 6.92 & 6.78--7.13 \\
\bottomrule
\end{tabular}

\end{table}

The $k=64$ rows also illustrate the overflow window of
Section~\ref{sec:query}.  At 64 lanes the coefficient bound is $9216$, so
overflow is possible from $E=115$ rather than $E=116$.  An encoding that
keeps the 32-lane window silently misses the $E=115$ case and reports a
boundary one step too high.  With the window derived from $k$, both 64-lane
boundaries match \eqref{eq:bhat}: 116 for the fold and 117 for the balanced
tree.

\subsection{Hardware conformance}
\label{sec:hardware}

Theorem~\ref{thm:faithful} is about separate binary32 operations with round
to nearest even and gradual underflow.  To check that this models something
real, we wrote the direct and hoisted forms as CUDA C++ kernels for an NVIDIA
RTX 5080, compiled with fused multiply-add disabled, and confirmed by
inspecting the generated PTX and SASS that the emitted code contains the
modeled multiplications and additions and no fused or flush-to-zero
instructions.  The 32-lane block of Section~\ref{sec:intro} replays bit for
bit: the left-fold direct kernel returns \texttt{0x7f800000}, and the hoisted
and balanced-tree kernels return \texttt{0x7f7b0000}.  Across 65\,536 random
blocks per tree with codes inside the guard, the direct and hoisted kernels
agree bitwise; across the same number outside the guard they differ on
99.8\% of blocks.  A $512\times512\times1024$ matrix product built from
guarded blocks matches bitwise across eight build variants.  These runs are a
consistency check on the model, not evidence for the theorem: no random draw
lands near $\Bhat_\rho(k)$, where the first divergence lives, which is
exactly why a proof is needed.

\section{Using the Guard in a Compiler}
\label{sec:compiler}

\subsection{The rewrite rule}

The compiler rule, in terms of the eight-bit scale fields, is
\begin{equation}
\frac{e_A\le252\qquad e_B\le252\qquad 107\le e_A+e_B\le254+\Bhat_\rho(k)}
     {\rD^\rho\;\rightsquigarrow\;\rH^\rho}
\label{eq:rule}
\end{equation}
where $\rD^\rho\rightsquigarrow\rH^\rho$ means that the direct form may be
replaced by the hoisted form, and the replacement preserves $\simeq$ on every
block whose codes satisfy the condition.  Since $\simeq$ is symmetric, the same
condition can be used to justify the reverse replacement, and an implementation may lower it to
per-lane dequantization.  At $k=32$ the condition is $107\le e_A+e_B\le371$
for a fold and $\le372$ for a balanced tree.

The solver certifies bounds for a specific format, block length, reduction tree and 
accumulator semantics. These choices can vary across kernel configurations,
like tiling, split-K and autotuning, meaning certification may not simply be once per format.
Our low-cost certification allows us to check the guard for each configuration efficiently,
allowing it to serve as a legality oracle for automated kernel generation and
rewrite search where many candidates need to be checked.
In our prototype, the bound $\Bhat_\rho(k)$ is computed offline,
when a lowering rule for a given block length and tree is written, and stored with the rule; the solver
is not on the compilation path.  At lowering time the condition is checked
in one of two ways.  With \emph{static range metadata}, a quantizer or
runtime states a guaranteed range for each operand's scale codes, range
propagation checks the condition once, and the compiler emits the hoisted form
with no branch.  Without static ranges, the compiler emits a \emph{runtime
check} of the four comparisons per block, or per group of blocks, taking the
hoisted form when it passes and the direct form otherwise.

This rule differs from attaching \texttt{reassoc}, \texttt{nnan}, and
\texttt{ninf} flags.  When the condition holds, it is guaranteed that the rewrite
preserves the value, including the exceptional values; if it doesn't the
compiler may still rewrite under an approximate policy, but cannot guarantee that the
result is value-preserving.

\subsection{An MLIR prototype}
\label{sec:mlir}

We implemented the rule in an MLIR~\cite{lattner2021mlir} prototype based on LLVM revision
\texttt{3f170b5b}. In our implementation, an operation \texttt{mxfp4.block\_dot} carries the
element type, the scale type, the block length, the reduction tree, and
optional scale-code ranges, which the guard needs and
which a generic lowering to scalar \texttt{arith} operations would discard.
The pass \texttt{mxfp4-lower-block-dot} emits target-independent scalar IR
without fast-math flags.  It uses the certified bounds for eleven
$(k,\rho)$ pairs, all fold lengths of Table~\ref{tab:boundaries} and balanced
trees for $k\in\{2,4,8,16,32\}$. Pairs without certified bounds use the direct form.
With static ranges the pass emits the hoisted form only
if both ranges lie inside $[0,252]$ and every possible code sum satisfies the guard.
With runtime checking it emits an \texttt{scf.if} whose condition checks the 4
comparisons on widened \texttt{i16} values and selects the hoisted form if they all pass,
and otherwise keeps the direct form with the requested tree.

Figure~\ref{fig:ir} shows the two lowerings on the block from
Section~\ref{sec:intro}.  On this block the direct form with a left fold returns \texttt{+inf}
and the hoisted form returns $3.3363623\times10^{38}$. 
Lit tests check the certified constants, runtime guard construction,
fallback behavior, and rejection of malformed metadata.

\begin{figure}[t]
  \centering
  \scriptsize
\begin{verbatim}
// direct form
%2 = arith.scaling_extf %0, %arg2 : f4E2M1FN, f8E8M0FNU to f32
%3 = arith.scaling_extf %1, %arg3 : f4E2M1FN, f8E8M0FNU to f32
%4 = arith.mulf %2, %3 : f32
// ... lanes 1--31, then 31 arith.addf in the requested tree ...
return %190 : f32

// hoisted form
%2 = arith.extf %0 : f4E2M1FN to f32
%3 = arith.extf %1 : f4E2M1FN to f32
%4 = arith.mulf %2, %3 : f32
// ... lanes 1--31, then 31 arith.addf in the requested tree ...
%191 = arith.extf %arg2 : f8E8M0FNU to f32
%192 = arith.extf %arg3 : f8E8M0FNU to f32
%193 = arith.mulf %191, %192 : f32
%194 = arith.mulf %193, %190 : f32
return %194 : f32
\end{verbatim}
  \caption{The two lowerings of one \texttt{mxfp4.block\_dot} as emitted by
    \texttt{mxfp4-opt}; lane 0 is shown and the other lanes are elided.  On
    the block of Section~\ref{sec:intro} with a left fold, the direct form
    returns \texttt{+inf} (\texttt{0x7f800000}) and the hoisted form returns
    $3.3363623\times10^{38}$ (\texttt{0x7f7b0000}).
    The guard rejects hoisting with a left fold reduction tree.}
  \label{fig:ir}
\end{figure}

At $k=32$, when static scale ranges fulfil the guard conditions,
the hoisted lowering emits 34 \texttt{arith.mulf} operations and
no branch, while the direct lowering emits 64 scale applications and 32 products.
Which lowering is faster or more efficient depends on the target.  On the RTX
5080, experiments with software-emulated binary32 MXFP4 matrix multiply show that the relative
performance of the two lowerings depends on the tile/decode configuration
(Appendix~\ref{app:rtx5080}).
These measurements motivate treating scale placement
as a target-dependent optimization choice.
The prototype demonstrates how certified guards permit the compiler
to choose between
the two forms while preserving the specified binary32 semantics.

\subsection{Beyond MXFP4}
\label{sec:beyond}

The construction of Section~\ref{sec:semantics} uses three structural
assumptions: element products lie on a common power-of-two grid with a bounded
integer coefficients, scales are powers of two, and the reduction tree is
fixed.  Coefficient bounds are the primary format-dependent property.
The state machine can be applied by representing elements
as $m2^{-f}$ with $|m|\le M$. The coefficient budget for $k$ lanes is $M^2k$,
with $E=t-2f$, provided that $M^2k<2^{24}$
and that the cap is adjusted so that $M2^{x-f}$ stays below $\Omega$.
Raw products and individually scaled operands must be exactly representable.
E2M1 exactness is enabled by its two significant bits and its narrow range:
its nonzero magnitudes range from $0.5$ to $6$, represented with $f=1$, yielding $M=12$.
The OCP formats MXFP6 (E2M3 and E3M2 elements) and MXINT8 also satisfy the budget at $k=32$.
The exactness argument does not extend to MXFP8:
E4M3 elements range over eighteen binades, from $2^{-9}$ to
$448$, so two lane products in one block can differ in scale by more than $2^{35}$,
and a sum of such products need not fit in 24 bits even though each product
alone does. 

NVFP4 uses an E4M3 scale instead of the power-of-two scale,
whose nontrivial significands can introduce rounding in subsequent operations.
Appendix~\ref{app:nvfp4} shows that, for positive finite E4M3 scales
and $k\le16$, every operation in both forms is exact in the single-level
model, so the rewrite is legal without a band.  Including the per-tensor binary32 scale, as
NVFP4 uses in practice, permits addition rounding at every node.
The present exactness argument does not cover this setting.
The implementation and certified bounds presented here cover MXFP4 with E2M1 elements and E8M0 scales.

\section{Related Work}

\paragraph{Microscaling.}
The OCP specification defines the MX formats and leaves the internal
precision and operation order of a dot product to the
implementation~\cite{ocp2023mx}; Rouhani et al.\ evaluate them for training
and inference~\cite{rouhani2023microscaling}.  Cuyckens et al.\ treat the
reduction tree of an MX datapath as a hardware design
parameter~\cite{cuyckens2026precision}; our results say that the parameter
also changes which rewrites are legal.  MX+ exploits the shared exponent to
detect outliers during serving~\cite{lee2025mxplus}.  Rout and Tine describe
a fused dot-product unit for mixed-precision tensor cores that scales
operands before the multiply-add~\cite{rout2025tenfour}, which is the direct
form.

\paragraph{Verifying floating-point optimizations.}
LifeJacket checks LLVM's floating-point peephole rules with bit-precise SMT
and found incorrect ones~\cite{notzli2016lifejacket}; Alive2 performs bounded
translation validation for LLVM IR~\cite{lopes2021alive2}.  Both reason
about scalar rules over arbitrary floating-point inputs and utilize floating-point theory.
We fix the element alphabet, scale format, and
tree, which allow bypassing the theory, and we produce a parametric region
rather than a decision on a single rule.
Icing provides a verified semantics for fast-math rewrites inside the CakeML
compiler~\cite{becker2019icing} - the programmer indicates which rewrites are allowed
through annotations. Our guard instead proves that the specific rewrite is value-preserving
on its domain.  Valpey et al.\ formalize the mixed-precision dot products of
three tensor-core generations in SMT~\cite{valpey2025tensor}; such bit-precise
hardware models are complementary to our work, supplying the accumulator semantics under
which a band would have to be re-derived (Section~\ref{sec:beyond}).
Volta checks the equivalence of GPU
kernels, including matrix multiplication and attention, over expressions in
the real numbers~\cite{dubey2025volta}; over reals the direct and
hoisted forms are equal by distributivity, and the witness of
Section~\ref{sec:intro} does not exist.

\paragraph{Floating-point decision procedures.}
Alternative routes include FP-to-bit-vector encoding~\cite{gadelha2020api},
abstract CDCL over floating-point intervals in
MathSAT~\cite{brain2014acdcl}, real-arithmetic axiomatization with
rounding lemmas~\cite{conchon2017threetier}, and a lossless reduction to
exact arithmetic~\cite{leeser2014real}, and translation of mixed real--floating-point formulas into
equisatisfiable real-arithmetic ones~\cite{salvia2019fprock}.
Our method trades their
generality for an exact linear-integer model of a restricted family.
Leeser et al. ignore subnormals and non-real results, whereas overflow,
NaN, and subnormal rounding are central to the rewrite studied here.
Gadelha et al.\ report performance differences across solver backends.
To control for this, we compare the floating-point and
integer encodings using the same solver.

\paragraph{Exact reasoning about floating-point networks.}
SELTZO verifies floating-point accumulation networks with a
precision-independent abstraction that reduces to integer
reasoning~\cite{zhang2025seltzo}; our reduction is in the same spirit but
exploits a finite alphabet and a shared exponent, and it is exact rather than
approximate on its domain.  FLoPS formalizes P3109 scalar formats and their
operations in Lean~\cite{chang2026flops}; it does not model MX blocks or
reductions.
Herbie and Daisy improve or bound rounding
error~\cite{panchekha2015herbie,darulova2018daisy}; our property is
equality, including of infinities and NaN, which error bounds do not express.
Fasi and Mikaitis emulate stochastic rounding on IEEE
hardware~\cite{fasi2021stochastic}; a stochastically rounded implementation
would need a distributional rather than a bitwise notion of equivalence.

\paragraph{Mixed precision and silent errors.}
Low-precision arithmetic has spread beyond ML into numerical linear
algebra~\cite{higham2022mixed} and is being adopted in scientific
codes~\cite{kashi2025mixed}, where results are checked against references
rather than validation scores.
Fleet-scale result checking for silent data corruptions~\cite{hochschild2021cores,dixit2021silent}
has been used for detecting rare, input-dependent,
and quiet divergences like the one in
Section~\ref{sec:intro}.  In contrast, a miscompilation is deterministic, and
its condition can be proved once rather than watched for unlike a hardware fault.

\section{Limitations}

The guarantees in this paper apply to a specific execution model:
each operation evaluated separately in binary32 with round
to nearest ties-to-even and gradual underflow.  Within the certified band every lane
product is exactly representable in binary32, so replacing the multiplication followed by addition
with a fused multiply-add removes no rounding and preserves the result.
However, wider accumulators, flush-to-zero, or a reduction tree other than the one
certified can change the behavior and require a separate equivalence argument.
In particular, the lower endpoint $-147$ relies on gradual underflow which would not hold under flush-to-zero.
In a full matrix multiplication, there is a cross-block accumulation, and the block-local
rewrite composes with it only if both implementations use the same outer accumulation order and arithmetic.
Theorems~\ref{thm:faithful}
and~\ref{thm:safe} have mathematical proofs supported by exhaustive and differential checks,
but haven't been mechanized. Formalizing the integer semantics and its connection to an
existing IEEE formalization is future work.  Sharpness is
established using the solver for the specific trees and lengths in
Tables~\ref{tab:boundaries} and~\ref{tab:scaling}.
The MLIR prototype generates reference scalar IR implementing the semantics.
The rewrite proof does not verify subsequent LLVM transformations or machine-code
generation. The prototype does not provide a production kernel.

\section{Conclusion}

An exact reduction to linear integer arithmetic makes a specific family
of floating-point equivalence queries tractable, while preserving the
behavior at range boundaries that determines legality. Implementing 
the reduction for MXFP4 helps derive a sufficient condition
for the legality of rewrites in the form of a cap on each scale code and 
a band for their sum, depending only on
the size of the largest proper subtree of the reduction tree.
Solver queries establish that the condition is also necessary for
several representative configurations. An MLIR prototype utilizes the results
in static or runtime checks consisting of a few simple conditions
to verify the legality of rewrites.
Retaining the format and reduction structure in the IR
enables checking legality conditions to guide lowering choices.

\bibliographystyle{splncs04}
\bibliography{refs}
\clearpage
\appendix
\renewcommand{\theHsection}{appendix.\Alph{section}}
\section{Single-Level NVFP4}
\label{app:nvfp4}

We consider E2M1 elements, positive nonzero finite E4M3 block scales,
$1\le k\le16$, and the separate binary32 operations of
\eqref{eq:direct} and~\eqref{eq:hoisted}.  Additional per-tensor scales
are not included.  Non-power-of-two scales raise the possibility of
rounding-induced divergence away from range boundaries.  The significand
bound below excludes this behavior in the single-level model.

We denote each E4M3 scale as $u2^v$ with integer $1\le u\le15$, and denote
each element as $m/2$ with $|m|\le12$.  Each individually scaled element
requires at most six significant bits, the scale product at most eight,
and each scaled lane product at most twelve.  For any subset of lanes,
the common-scale coefficient is bounded by
\[
  u_Au_B\left|\sum_i m_{a,i}m_{b,i}\right|
  \le225\cdot144k\le518\,400<2^{24}.
\]
Therefore, every partial sum fits within binary32's 24-bit significand.
The scale range $[2^{-9},448]$ places every nonzero scaled subset sum
between $2^{-20}$ and $16\cdot36\cdot448^2<2^{27}$ in magnitude;
the other nonzero intermediates are also normal and finite.  Thus all
operations are exact, and both forms return the exact real dot product
for any fixed binary tree in this range.  Positive scales preserve the
signs of zero, and corresponding exact additions follow the same zero-sign
rules, so the forms also agree bitwise.

This was validated through enumeration of all
$16^2\cdot126^2=4\,064\,256$ one-lane cases and replay of 200\,000 random
blocks for $k\in\{2,4,16\}$.  No divergences or deviations
from the exact real result were observed.  The direct SMT query returned
\unsat{} at $k=1$;
at $k=2$, \unsat{} results for the multiplication and sum-distributivity
lemmas establish equivalence by composition.  For larger block lengths,
the SMT queries exceeded the five-minute time limit.  The result through
$k=16$ therefore follows from the preceding argument, with replay
providing supporting validation.

Scale hoisting is therefore valid throughout this single-level domain.  MXFP4's divergences arise at range
boundaries that remain unreachable with E4M3 scales in this setting.
The result does not cover full NVFP4 with an additional per-tensor scale.

\section{RTX 5080 Scale-Placement Measurements}
\label{app:rtx5080}

Table~\ref{tab:rtx5080-tiles} reports the complete twelve-configuration sweep
of a software-emulated MXFP4 matrix multiplication with
$(M,N,K)=(4096,2560,2560)$ and 32-lane blocks.  The hand-written CUDA
kernels implement separate binary32 operations with round-to-nearest-even,
gradual underflow, a left-fold block reduction, and binary32 output.
They implement the modeled arithmetic but are not generated by the MLIR
prototype.  These measurements concern software emulation, not native
block-scaled tensor-core instructions.

The recorded environment is an NVIDIA GeForce RTX 5080 (16\,GiB,
\texttt{sm\_120}), driver 590.48.01, NVRTC and \texttt{ptxas} 12.8,
and PyTorch 2.11.0+cu128.  Both compilation stages disable FMA contraction;
the kernels explicitly use rounded binary32 multiply and add instructions.
Inputs are Gaussian tensors generated with seed 5 and quantized once;
quantization and compilation are outside the timed region.  Each entry is
the median of 15 CUDA-event measurements after five warmups.  Configurations
are visited in the table's order, with direct measured before hoisted at
each configuration.  The historical log retains medians, not individual
timing samples, so no confidence intervals are inferred from this sweep.

\begin{table}[t]
  \centering
  \caption{Recorded RTX 5080 kernel times.  Micro-tile gives the output
    elements computed per thread; thread block gives the two-dimensional
    thread arrangement.  ALU denotes bit-assembly decoding and LUT a
    shared-memory lookup table.  $\Delta=100(T_D-T_H)/T_D$ is the percentage
    reduction in elapsed time; negative values indicate a slowdown.
    Every configuration reports zero bitwise mismatches between the forms
    and against the first configuration's direct output.}
  \label{tab:rtx5080-tiles}
  \small
  % Generated by analyze_gpu.py from the recorded ablation_tile.json.
\begin{tabular}{@{}llcrrr@{}}
\toprule
Micro-tile & Thread block & Decode & $T_D$ (ms) & $T_H$ (ms) & $\Delta$ (\%) \\
\midrule
$2\times2$ & $16\times16$ & ALU & 24.342 & 24.180 & $+0.67$ \\
$2\times2$ & $16\times16$ & LUT & 27.392 & 22.669 & $+17.24$ \\
$4\times4$ & $16\times16$ & ALU & 13.720 & 14.070 & $-2.56$ \\
$4\times4$ & $16\times16$ & LUT & 12.074 & 11.151 & $+7.64$ \\
$4\times4$ & $8\times32$ & ALU & 13.559 & 14.340 & $-5.76$ \\
$4\times4$ & $8\times32$ & LUT & 11.765 & 11.447 & $+2.71$ \\
$4\times8$ & $16\times8$ & ALU & 12.015 & 12.941 & $-7.71$ \\
$4\times8$ & $16\times8$ & LUT & 15.960 & 15.240 & $+4.51$ \\
$8\times4$ & $8\times16$ & ALU & 13.471 & 13.969 & $-3.69$ \\
$8\times4$ & $8\times16$ & LUT & 15.645 & 13.892 & $+11.20$ \\
$8\times8$ & $8\times16$ & ALU & 10.361 & 11.460 & $-10.61$ \\
$8\times8$ & $8\times16$ & LUT & 18.374 & 13.798 & $+24.91$ \\
\bottomrule
\end{tabular}

\end{table}

Hoisting reduces the measured time at all six LUT configurations and at
one of the six ALU configurations.  Configuration choice dominates the
best observed result: the fastest direct configuration takes 10.361\,ms,
whereas the fastest hoisted configuration takes 11.151\,ms, or 7.6\% longer.
This finite sweep establishes neither a global optimum nor a general
speedup.  It supports considering scale placement alongside other compiler
choices, with certified guards establishing when a candidate rewrite is legal.
The experiment compares direct and statically hoisted forms; it does not
measure the cost of runtime guard evaluation.

The artifact includes the original median log, environment record, benchmark
source, and a CPU-only table-regeneration script.  Its optional GPU command
reruns all twelve configurations on a separately configured CUDA system;
the CPU reproduction path checks the recorded data and regenerates the
table without claiming a new GPU measurement.
\end{document}